\documentstyle[aps,prb,preprint]{revtex}
\title{Normal state resistivity, upper critical field and Hall effect of superconducting perovskite $MgCNi_3$}
\author{S. Y. Li, R. Fan, X. H. Chen$^{\ast}$, C. H. Wang, W. Q. Mo, K. Q. Ruan, Y. M. Xiong, X. G. Luo, H. T. Zhang, L. Li, Z. Sun, and L. Z. Cao}
\address{ Structural Research Laboratory and Department of Physics,
University of Science and Technology of China,  Hefei, Anhui 230026,
P. R. China}
\date{\today}
\begin{document}

\maketitle

\begin{abstract}

The normal state resistivtity, upper critical field $H_{c2}$ and
Hall coefficient $R_H$ in superconducting perovskite $MgCNi_3$
($T_c \approx 8 K$) have been studied. Above 70 K, $\rho(T)$ fits
well curve predicted by Bloch-Gr\"{u}neisen theory consistently
with electron-phonon scattering. $H_{c2}(0)$ was estimated to be
about 15.0 Tesla within the weak-coupling BCS theory, and the
superconducting coherence length $\xi(0)$ is approximately 47 \AA.
$R_H$ of $MgCNi_3$ is negative for the whole temperature range
which definitely indicates that the carrier in $MgCNi_3$ is
electron-type. $R_H$ is temperature independent between $T_c$ and
$\sim$ 140 K. Above $\sim$ 140 K, the magnitude of $R_H$ decreases
as temperature rises. At T = 100 K,  the carrier density is $1.0
\times 10^{22}/cm^3$, which is comparable with that in perovskite
$(Ba,K)BiO_3$, and less than that of the metallic binary $MgB_2$.

\end{abstract}

\vskip 10 pt


\section*{introduction}

The recent discovery of superconductivity in the simple
intermetallic compound $MgB_2$\cite{Nagamatsu} has attracted great
attention because of its relatively high transition temperature
($T_c$ = 39 K) and the highly promising potential application. It
suggests that intermetallic compounds with simple structure types
are worth serious reconsideration as sources of new
superconducting materials. More recently, the observation of
superconductivity at 8 K in the perovskite structure intermetallic
compound $MgCNi_3$\cite{Cava} indicates that $MgB_2$ will not be
the only one of its kind within the chemical paradigm for new
superconducting materials.

The variable stoichiometry compound $MgC_xNi_3$, for $0.5 < x <
1.25$, has been reported and assigned to the perovskite structure
type by analogy in 1950's\cite{Scheil,Huetter}, but neither its
crystal structure nor its physical properties had been determined.
By using powder neutron diffraction, He et al.\cite{Cava} has
found that the superconducting phase in nominal composition
$MgC_{1.25}Ni_3$ is $MgC_{0.96}Ni_3$ with the classical cubic
perovskite structure, space group $Pm-3m$. They also determined
the electron-phonon coupling constant $\lambda_{ph} \sim 0.77$ by
specific heat measurements. Although this value of $\lambda_{ph}$
is in the range of conventional phonon, more properties of both
normal state and superconducting state need to be clarified to
determine the microscopic mechanism of superconductivity in this
compound. A complete structural and electronic equivalence of the
superconducting oxide perovskites like $(Ba,K)BiO_3$
(BKBO)\cite{Taraphder} and intermetallic perovskite superconductor
$MgCNi_3$ has been considered\cite{Cava}. For the oxide
perovskites, an important characteristic of the superconductivity
is that the electronic states at the Fermi energy involve holes in
the oxygen electronic orbitals. Preliminary band structure
calculations have shown that the electron states at the Fermi
surface of $MgCNi_3$ are dominated by the $3d$ orbitals of
Ni\cite{Cava2}, so the conduction may also involve holes in Ni
$d$-states. Although it is expected that the carriers should be
holes as monovalent potassium replaces divalent barium in BKBO,
the Hall coefficient measurement\cite{Sato} on BKBO thin films
indicates that the charge carriers are electrons. Therefore the
measurement of the Hall effect in $MgCNi_3$ should be very
interesting.

In this paper, we report the study of the normal state resistivity and the first measurements of the upper critical field $H_{c2}$ and Hall coefficient $R_H$ for superconducting perovskite $MgCNi_3$. It is found that at high temperature (above 70 K) $\rho(T)$ can be fitted well by Bloch-Gr\"{u}neisen theory. $H_{c2}-T$ phase diagram was obtained. The sign and temperature dependence of $R_H$ were determined by Hall effect measurement, which definitely indicates that the carrier in $MgCNi_3$ is electron-type.

\section*{experiment}

To obtain perfectly stoichiometric composition $MgCNi_3$ with
highest $T_c$, excess carbon is required\cite{Cava}. Due to the
volatility of Mg during the synthesis of this compound, excess Mg
is needed. In this study, sample with nominal formula
$Mg_{1.2}C_{1.4}Ni_3$ was prepared. Starting materials were bright
Mg flakes, fine powders of Ni, and the powders of amorphous carbon
with high purity. Starting materials were mixed, ground for a few
minutes, and pressed into pellet. The pellet was loaded in a Ta
foil, which was in turn sealed in a stainless steel reactor. The
above operations were carried out in a glove box with an Ar
environment, in which the content of oxygen and water is less than
a few ppm. The reactor was fired in a tube furnace under high-pure
Ar atmosphere for half an hour at 600$^o$C, followed by an hour at
900$^o$C. After cooling, the sample was reground, pressed into
pellet, and sintered for another one hour at 900$^o$C under the
same environment. The resulting sample was dense with a length of
7 mm and a width of 2 mm. The structure was characterized by
powder x-ray diffraction (XRD) analysis using Rigaku
D/max-$\gamma$A x-ray diffractometer (XRD) with graphite
monochromatized Cu K$\alpha$ radiation ($\lambda$ = 1.5406 \AA).
The data were collected over 2$\theta$ range from 15$^o$ to 75$^o$
with a 0.02$^o$ step.

In order to obtain a good Hall voltage signal, the sample was
mechanically polished until it was very thin (360 $\mu$m). The
longitudinal and Hall voltages were measured by using the standard
dc 6-probe method. The magnetic field was applied perpendicular to
the sample surface by using a superconducting magnet system
(Oxford Instruments) and the applied current is 15 mA. The Hall
voltage was extracted from the antisymmetric parts of the
transverse voltages measured under opposite directions to remove
the longitudinal component due to the misalignment of the Hall
voltage pads. The Hall voltage was found to be linear in the
magnetic field.

\section*{results and discussion}

Fig. 1 shows temperature dependence of resistivity under zero
field and the XRD pattern (inset) for the sample with nominal
composition $Mg_{1.2}C_{1.4}Ni_3$. XRD pattern indicates that the
sample is nearly single phase. By the least-square fitting to the
positions of 7 Bragg reflection peaks between 2$\theta$ values of
15$^o$ and 75$^o$, the cubic cell parameter $a$ = 3.81154(5) \AA\
is obtained, being consistent with previous report\cite{Cava}. The
resistive superconducting transition is very sharp. The midpoint
of the resistive transition is 8.0 K, the 90-10\% transition width
is less than 0.3 K, and the resistive onset temperature is 8.2 K.
$T_c$ of the sample is nearly the same as that reported in Ref. 2.
The resistivity ratio $\rho_{300K}/\rho_{9K}$ is 2.5 which is
larger than that in Ref. 2. Also the normal state resistivity is about 3
times larger than that in Ref. 2. It is important to note that the
shape of $\rho(T)$ curve is almost identical to that reported by
He et al.\cite{Cava}. One can observe an upward curvature of
$\rho(T)$ curve at low temperature followed by a downward one for
higher temperature. It looks like that of conventional metals and
different from the simply linear-T dependence observed in  copper
oxide superconductors. The similar shape of $\rho(T)$ curve has
been observed in BKBO thin film\cite{Moon} and single
crystal\cite{Affronte} in which $\rho(T)$ fits well curve
predicted by Bloch-Gr\"{u}neisen theory consistently with
electron-phonon scattering\cite{Ziman}.

In order to investigate to what extent the resistive behavior is consistent with electron-phonon scattering mechanism, we fit the resistivity data with an explicit form of the Bloch-Gr\"{u}neisen expression which is valid in the case of an Einstein phonon distribution $\epsilon=k_B\Theta_E$\cite{Engquist} as in Ref. 9:
\begin{equation}
   \rho^{-1} = \rho_{p}^{-1} + (\rho_0 + \rho_{ph})^{-1}
\end{equation}
\begin{equation}
   \rho_{ph} = \rho_lcoth(\Theta_E/2T)[1+(2/3)sinh^2(\Theta_E/2T)]^{-1}
\end{equation}
$\rho_0$ and $\rho_p$ are respectively a residual and a parallel
resistivity, respectively. $\rho_l$ is a constant. It is found
that between 70 K and 300 K $\rho(T)$ fits well curve by the
Bloch-Gr\"{u}neisen expression. The solid line in Fig.2 represents
the result of the best fit to the experimental data. The best fit
result gives $\Theta_E$ = 206(1) K, $\rho_0$ = 202(1)
$\mu\Omega$cm, $\rho_p$ = 574(1) $\mu\Omega$cm, and $\rho_l$ =
177(1) $\mu\Omega$cm. The Einstein temperature $\Theta_E$ is
comparable with that of BKBO\cite{Affronte}. Below 70 K, the
fitting is not well and gives very low $\Theta_E$. We speculate it
that the Einstein phonon distribution is not proper and continuous
phonon spectra, such as Debye spectrum, should be adopted at low
temperature. We simply fit the low temperature resistivity data
using a power law, $\rho = \rho_0 + aT^n$. The fit result gives
$\rho_0$ = 120.7(1) $\mu\Omega$cm, n = 1.46(1). The fit curve is
also shown as an unbroken line in Fig.2.

Fig. 3 shows the $\rho(T)$ curves under magnetic fields up to 14
T. The resistive superconducting transition shifts to low
temperature with increase of the magnetic field. The onset
temperature decreases and the broadening of the superconducting
transition is almost absent with the increasing magnetic field.
This result is in contrast to the high-$T_c$ cuprate
superconductors and similar to A15 intermetallic compounds such as
$Nb_3Sn$. Fig. 4 shows the $H_{c2}-T$ phase diagram obtained from
the $\rho$ vs. $T$ curves at different fields. Here $T_c$ is
defined as the intersection of the linear extrapolation of the
most rapidly changing part of $\rho(T)$ and that of the normal
state resistivity, as shown in the upper inset of Fig. 4. The
lower inset of Fig. 4 is the magnetic field dependence of the
resistance of the sample at 2.3 K. Defining $H_{c2}$ as the
intersection of the linear extrapolation of the most rapidly
changing part of $\rho(H)$ and that of the normal state
resistivity, one can get $H_{c2}(2.3K)$ = 12.5 T, which is in good
agreement with the result from temperature dependence $\rho(T)$.
Within the weak-coupling BCS theory, $H_{c2}(T=0)$ can be
estimated using the Werthamer-Halfand-Hohenberg(WHH)
formula\cite{WHH},
\begin{equation}
   \mu_0H_{c2} = -0.693(dH_{c2}/dT)_{T=T_c}T_c
\end{equation}
which leads to a $\mu_0H_{c2}$ value of 15.0 T. Meanwhile the Pauli-limiting field
\begin{equation}
   \mu_0H^{Pauli} = 1.24K_BT_c/\mu_B
\end{equation}
expected within the same weak-coupling BCS theory\cite{Clogston}
is also 15.0 Tesla for $T_c$ = 8.12 K, inferred from Fig. 4. One
can see that $\mu_0H^{WHH}$ and $\mu_0H^{Pauli}$ agree perfectly
with each other. The superconducting coherence $\xi(0)$ is
estimated to be approximately 47 \AA, using the Ginzburg-Landau
formula for an isotropic three-dimensional superconductor,
$\mu_0H_{c2} = \Phi_0/2\pi\xi^2(0)$.

Fig. 5 shows the Hall coefficient $R_H$ from room temperature down
to $T_c$ under 10 T. The two curves in the inset represent the
Hall voltage measured at 100 K for opposite magnetic fields up to
10 T. Clearly, the field dependent Hall voltage is symmetric and
linear. The $R_H$ is negative for the whole temperature range. We
found that $R_H$ is essentially temperature independent between
$T_c$ and $\sim$ 140 K, at least within our experimental accuracy.
Above $\sim$ 140 K, the magnitude of $R_H$ decreases as
temperature increases. It is found that $R_H(140K)/R_H(295K)$ is
about 1.3. At T = 100 K, $R_H$ = $-6.1 \times 10^{-10} m^3/C$, and
the calculated carrier density is $1.0 \times 10^{22}/cm^3$. The
main feature of the Hall effect is the negative $R_H$. It
definitely indicates that the carrier in $MgCNi_3$ is
electron-type, which is strongly supported by the negative thermoelectrical power observed from room temperature to 10 K\cite{Li}.
The carrier density is comparable with that of BKBO\cite{Sato},
and less than that of metallic binary $MgB_2$\cite{Kang} and
larger than that in copper oxide superconductors\cite{Harris}.
Another feature of the Hall coefficient is the fact that it
depends on temperature. For conventional isotropic metal with
ordinary electron-phonon scattering mechanism, the Hall
coefficient is expected to be temperature-independent. We note
that $R_{H}$ temperature behavior of $MgCNi_3$ is very similar to
that of $Ba_{1-x}K_{x}BiO_{3}$ single crystal\cite{Affronte}. A
good example of temperature dependent $R_H$ is copper based
superconducting oxides, in which $R_H(T)$ is explained by exotic
magnetic excitation\cite{Anderson}. Preliminary band structure
calculations have shown that the electron states at the Fermi
surface of $MgCNi_3$ are dominated by the $3d$ orbitals of
Ni\cite{Cava2}. The itinerant electrons arise from the partial
filling of the nickel $d$-states, which generally leads to
ferromagnetism as is the case in metallic. The temperature
dependence of $R_H$ may also be ascribed to magnetic excitations
as copper based superconducting oxides. However,
$Ba_{1-x}K_{x}BiO_{3}$ has no magnetic ion and manifests similar
temperature dependence of $R_H$ as $MgCNi_3$, so the temperature
behavior of $R_H$ remains an open question.

\section*{conclusion}

In summary, we have measured the temperature dependence of
resistivity, upper critical field and Hall effect for
superconducting perovskite $MgCNi_3$. Above 70 K, the normal state
$\rho(T)$ behavior follows Bloch-Gr\"{u}neisen theory consistently
with electron-phonon scattering, being similar to that of the
three-dimensional BKBO with the same structure. From $H_{c2}-T$
phase diagram, $H_{c2}(0)$ was estimated about 15.0 Telsa within
the weak-coupling BCS theory, the superconducting coherence length
$\xi(0)$ was found to be approximately 47 \AA. Negative $R_H$
definetly suggests that the carrier is electron character, is the
same as that in BKBO and different from that in $MgB_2$. At T =
100 K, $R_H$ = $-6.1 \times 10^{-10} m^3/C$, and the calculated
carrier density is $1.0 \times 10^{22}/cm^3$, which is comparable
with that in BKBO, and less than that of $MgB_2$ and larger than
that in copper oxide superconductors. $R_H$ is temperature
independent between $T_c$ and $\sim$ 140 K. Above $\sim$ 140 K,
the magnitude of $R_H$ decreases as temperature increases. The
temperature behavior of $R_H$ remains an open question.

\section*{acknowledgement}

This work is supported by the Natural Science Foundation of China and by the Ministry of Science and Technology of China (NKBRSF-G19990646).

\vskip 10 pt

\noindent $\ast$ To whom all correspondence should be addressed. Email: chenxh@ustc.edu.cn

\newpage
\noindent
{\bf FIGURE CAPTIONS} \\

\noindent
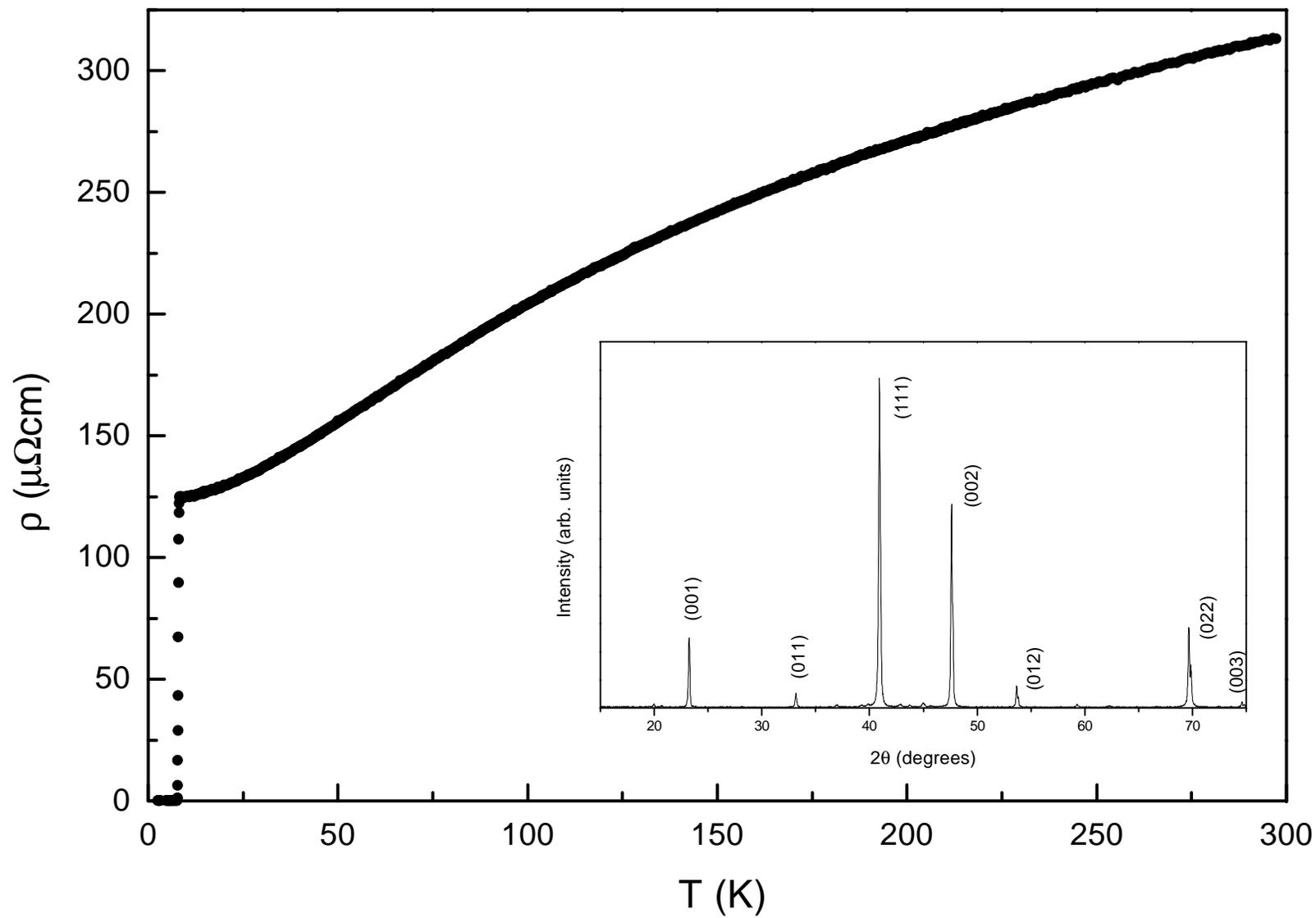
FIG 1:

The temperature dependence of resistivity for $MgCNi_3$ sample.
Inset: the XRD pattern.\\

\noindent
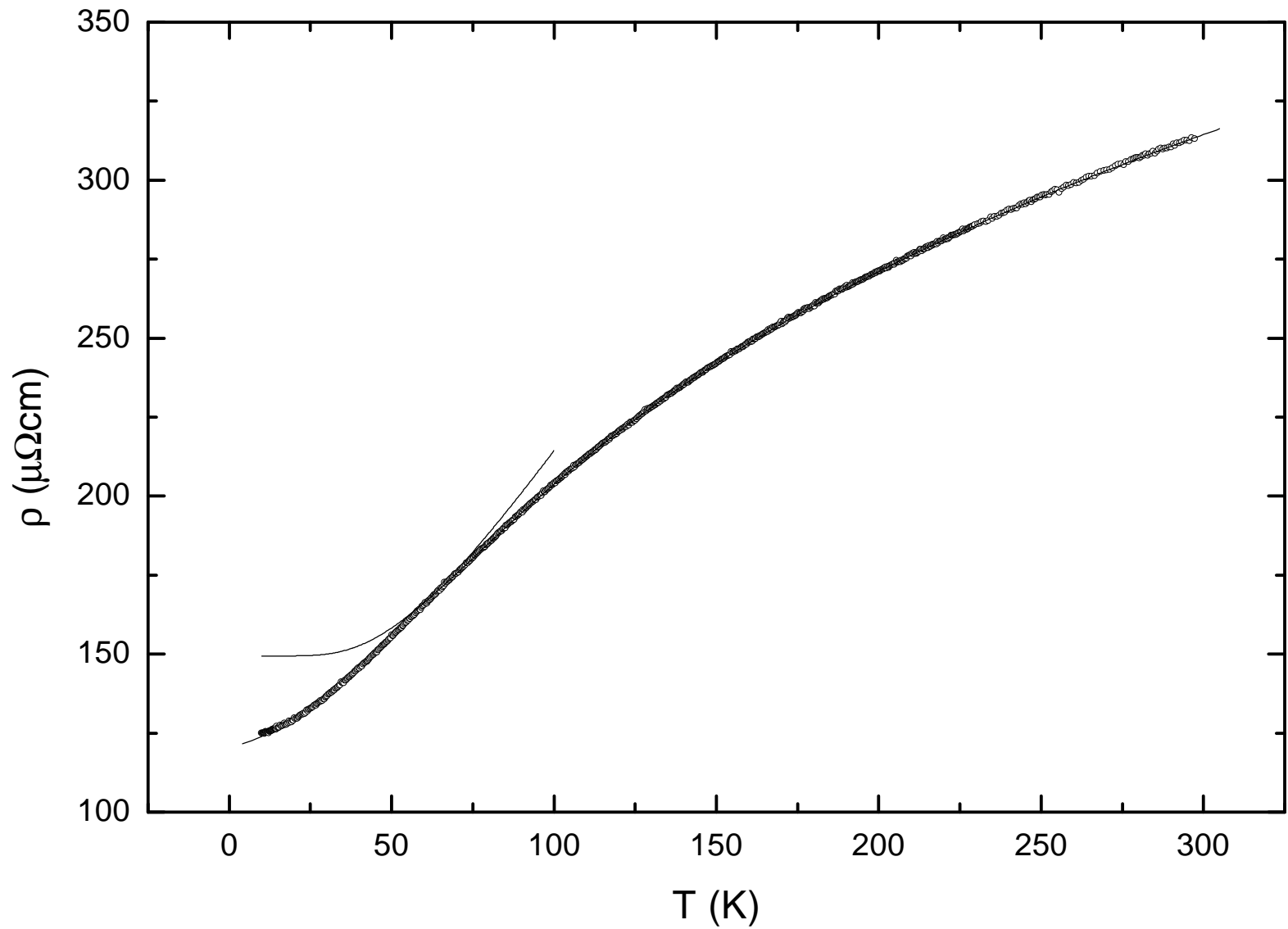
FIG 2:

The temperature dependence of normal state resistivity  for
$MgCNi_3$ sample. The solid lines are fit of $\rho(T)$ to
Bloch-Gr\"{u}neisen expression and power law, respectively.\\

\noindent
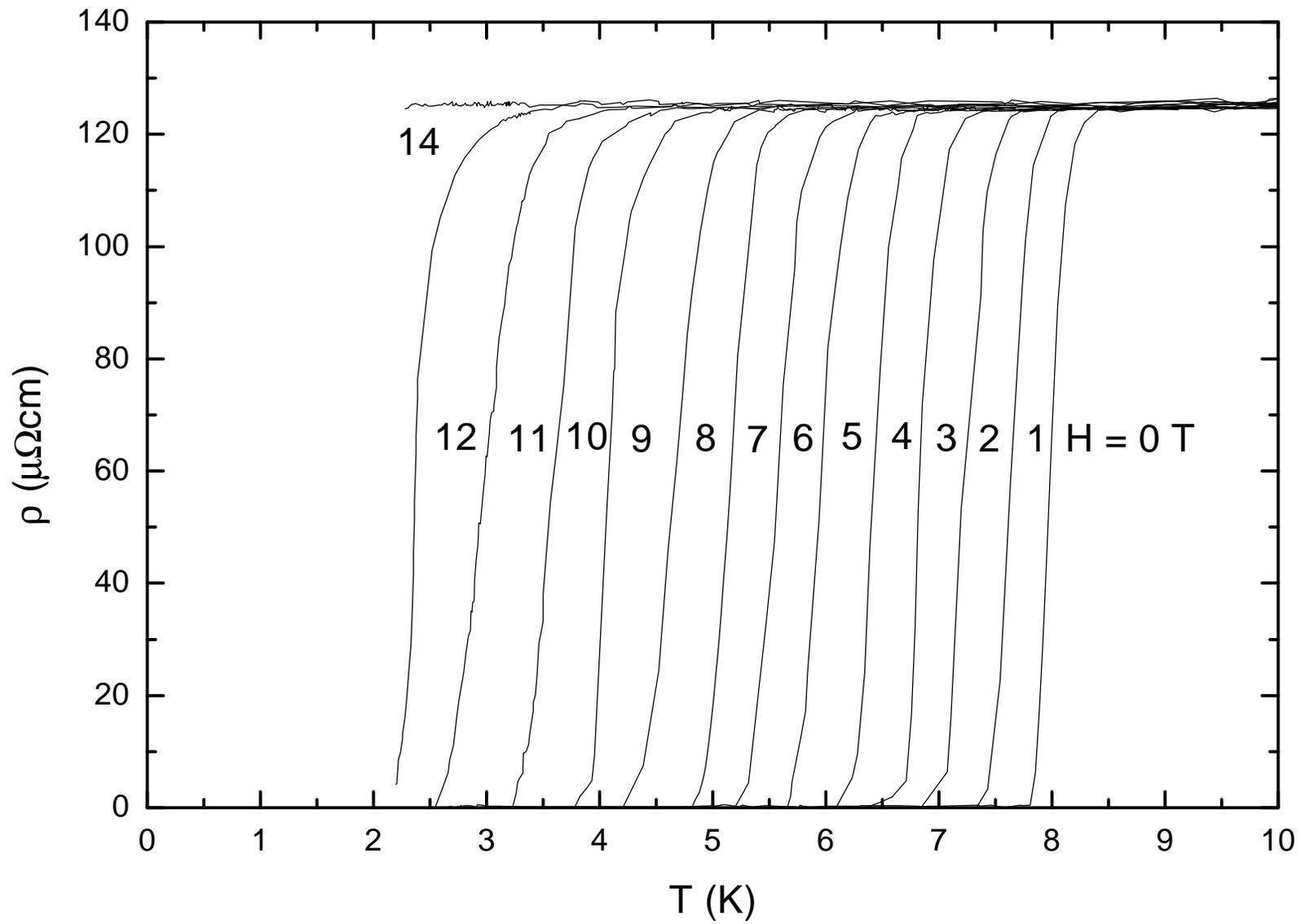
FIG 3:

The $\rho(T)$ curves under magnetic fields up to 14 Tesla for
$MgCNi_3$.\\

\noindent
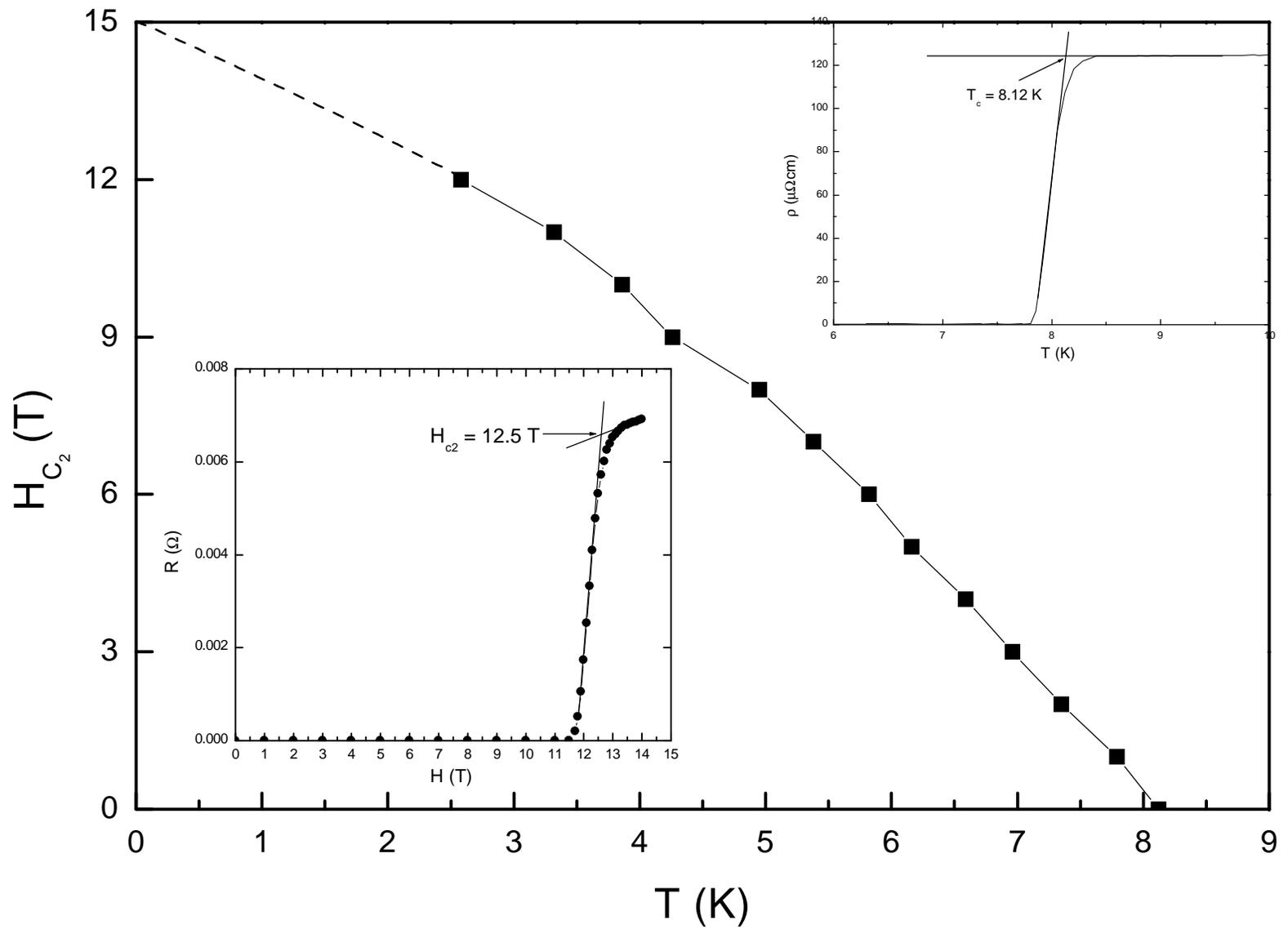
FIG 4:

Upper critical field $H_{c2}$ as a function of temperature for $MgCNi_3$. The up
per
inset shows the temperature dependence of sample resistivity under zero magnetic
 field.
The lower inset shows the magnetic field dependence of resistance at T = 2.3 K.\\

\noindent
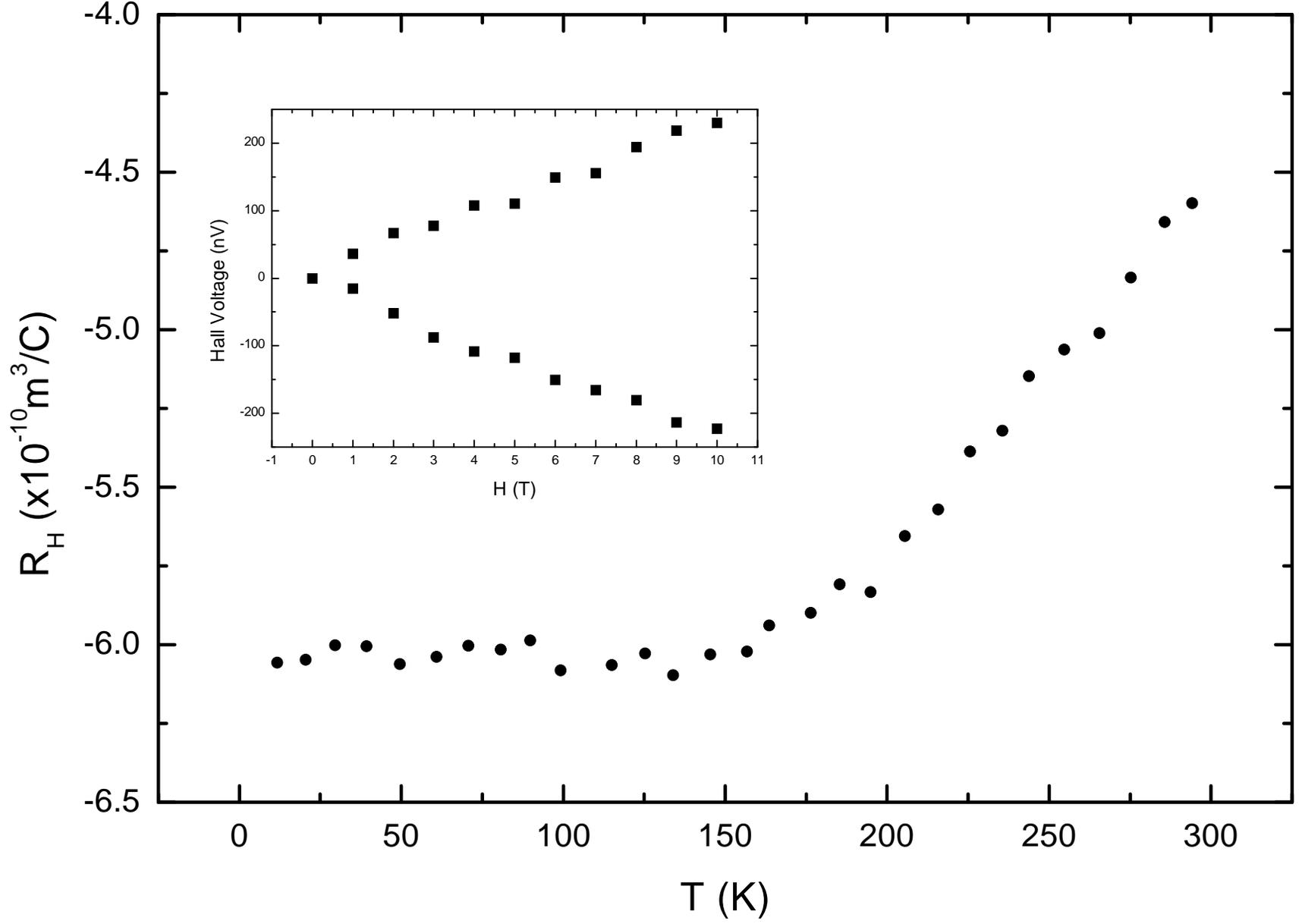
FIG 5:

Temperature dependence of Hall coefficient measured under 10 Tesla
for $MgCNi_3$. The two lines in the inset represent the Hall
voltage measured at 100 K for opposite two directions of the
applied field up to 10 T.\\

\end{document}